\documentstyle[11pt,epsfig]{article}
\def\a{\alpha}
\def\t{\theta}
\def\dis{\displaystyle}
\def\le{\left(}
\def\ri{\right)}

\def\no{\nonumber}

\def\rar{\rightarrow}

\def\e{\epsilon}
\def\f12{\frac{1}{2}}

\def\pd{\partial}

\def\L{\lambda}
\newcommand{\Li}{\mathop{\mathrm{Li}}\nolimits}

\begin{document}
\begin{titlepage}
\flushright{USM-TH-182}
\vskip 2cm
\begin{center}
{\Large \bf Towards the two-loop $Lcc$ vertex in Landau gauge} \\
\vskip 1cm  
Gorazd Cveti\v c$^{a}$, Igor Kondrashuk$^{a,b}$, 
Anatoly Kotikov$^{a,c}$, and 
Ivan Schmidt$^{a}$ \\
\vskip 5mm  
{\it  (a) Departamento de F\'\i sica, Universidad T\'ecnica 
Federico Santa Mar\'\i a, \\
 Avenida Espa\~{n}a 1680, Casilla 110-V, Valparaiso, Chile} \\
{\it  (b) Departamento de Ciencias Basicas, 
Universidad del Bio-Bio, Campus 
Fernando May, Casilla 447, Avenida Andreas Bello, 
Chillan, Chile} \\
{\it  (c) Bogoliubov Laboratory of Theoretical Physics, 
Joint Institute for 
Nuclear Research, \\
Dubna, Russia} \\
\end{center}
\vskip 20mm
\begin{abstract}
We are interested in the structure of the $Lcc$ vertex in the Yang-Mills theory,
where $c$ is the ghost field and $L$ the corresponding BRST auxiliary field.
This vertex can give us information on other vertices, and the possible
conformal structure of the theory should be reflected in the structure of
this vertex. There are five two-loop contributions to the $Lcc$ vertex in the 
Yang-Mills theory. We present here calculation of the first of the five contributions. 
The calculation has been performed in the position space. 
One main feature of the result is that it does not depend on any scale, 
ultraviolet or infrared. The result is expressed in terms of logarithms and Davydychev integral $J(1,1,1)$ that 
are functions of the ratios of the intervals between points of effective fields 
in the position space. To perform the calculation we apply 
Gegenbauer polynomial technique and uniqueness method. 
\vskip 1cm
\noindent Keywords: Gegenbauer technique 
\end{abstract}
\end{titlepage}

\section{Introduction}

Recently it has been shown that the effective action of the ${\cal N}=4$ SYM written in terms of the dressed mean fields does not depend on any scale, ultraviolet or infrared
\cite{Cvetic:2004kx,Cvetic:2006kk}. The theory in terms of these variables is invariant conformally. Therefore, investigation of this action should be greatly simplified.
It might be possible to fix all three-point correlation functions up to some coefficient. For example, the three-point function of dressed gluons in the Landau gauge 
could be found in such a way. Conformal symmetry does not help a lot in the case of the four-point function since an arbitrariness arises. 
However, in any case, it is important to check all these statements directly by the precise calculations of the vertices in terms of 
the dressed mean fields. $Lcc$ vertex is the most simple object for this calculation, especially in the Landau gauge. In that gauge it is simply 
totally finite. This fact has been indicated first in Refs.~\cite{Cvetic:2004kx,Kondrashuk:2004pu}.

The purpose of the study is to calculate the $Lcc$ vertex at two-loop level for  ${\cal N}=4$ super-Yang--Mills theory, which should confirm the statement of 
Refs.~\cite{Cvetic:2004kx,Cvetic:2006kk} that were derived from the results of Refs. \cite{Cvetic:2002dx,Cvetic:2002in,Kondrashuk:2000br,Kondrashuk:2003tw,Kondrashuk:2000qb}. 
The present paper contains the calculation of the first needed diagram,
and also all needed formulas. The evaluation of other diagrams is a subject of our future investigation. In our opinion, it is the first calculation of two-loop three-point 
diagrams in general kinematics for a real physical model. The calculation is very nontrivial and can be used by others in some different studies.

With a little modification, the investigation can be applied also for the finite (and of course, singular part) of the vertex at two-loop level for nonsupersymmetric theory. 
As to the singular part in higher orders for nonsupersymmetric case, in Ref. \cite{Dudal:2003pe} the explicit three-loop computation of the anomalous dimension of the operator 
$cc$ has been carried out. The vertex $Lcc$ is convergent superficially in Landau gauge in any gauge theory, supersymmetric or nonsupersymmetric. This fact is a consequence of 
the possibility to integrate two derivatives by parts and put them outside the diagram on the external ghost legs due to transversality of the gauge propagator. 
If we make a change of the normalization $L$ to $Lg,$ as it takes place in Refs. \cite{Dudal:2003pe,Dudal:2003np},  than the superficial 
divergence  would mean that $Z_L Z_g Z_c = 1$. (Renormalization constant of $g$ is $Z_g,$ and the renormalization constant of $c$ is $Z_c^{1/2}).$ This is the condition of superficial 
convergence of this vertex. Formally, this result holds to all orders of the perturbation theory due to the so-called Landau ghost equation, stemming from the fact that in the Landau gauge 
the Yang-Mills action is left invariant by a constant shift of the Faddeev-Popov ghost $c$ (Ref. \cite{Blasi:1990xz}). Note that it is not like the normalization used in 
Refs. \cite{Cvetic:2004kx,Cvetic:2006kk} where $Z_L Z_c = 1.$ Moreover, the renormalization of ghost and antighost fields in our Refs.  \cite{Cvetic:2004kx,Cvetic:2006kk} 
was chosen to be independent. This is different from Ref.\cite{Dudal:2003pe} in which $Z_c = Z_{\bar{c}}.$ Thus, in terms of conventions of Ref. \cite{Cvetic:2004kx,Cvetic:2006kk} 
$Z_L = Z_c = 1.$ It means the field $c$ does not get the renormalization in our convention. This coincides with the old results obtained from the antighost equation of Ref. 
\cite{Blasi:1990xz}. Neither does the external field $L$ get the renormalization in this convention. Starting with the two loop order infinities reproduce the renormalization of 
the gauge coupling in nonsupersymmetric theory. If we change the convention of the renormalization, the relation between the renormalization constants of paper \cite{Dudal:2003pe} 
for nonsupersymmetric case can be reproduced.

Knowing the structure of the $Lcc$ vertex, one can obtain other vertices in terms of this one by using Slavnov-Taylor identity \cite{Slavnov:1972fg,Taylor:1971ff,Slavnov:1974dg,
Faddeev:1980be,Lee:1973hb,Zinn-Justin:1974mc} which is a consequence of the BRST symmetry \cite{Becchi:1974md,Tyutin:1975qk}. Moreover, the algorithm for obtaining these structures is expected 
to be simple, due to the simple structure of the $Lcc$ vertex, in particular due to its scale independence. Similar arguments can be applied to ${\cal N}=8$ supergravity 
\cite{Bjerrum-Bohr:2006yw,Kang:2004cs}, and to other theories which possess high level supersymmetry to garantee good properties of the correlators. 
In some theories, for example Chern-Simons field theory, nonrenormalization of gauge coupling is protected by topological reasons and similar approach is valid  
near the fixed points \cite{Avdeev:1992jt} in the coupling space.

The article has the following structure. Section 2 contains the derivation of 
gluon and ghost propagators 
in the position space. A review of the one-loop results and all two-loop 
diagrams, contributing to the problem, are given in Section 3. Section 4 
demonstrates a useful representation for the two-loop diagram $(a)$ which is the 
subject of this study. In Section 5 we show basic formulas for calculation 
of the considered Feynman integrals. 
The calculation of diagram (a) is performed in Sections 6 and 7. Moreover,
details of the calculations can be found in  Appendices A and D.
The most complicated Feynman integrals are 
evaluated in Appendices B and C. 
Section 8 contains conclusions and a summary of the results, and 
discussion about the future steps.

\section{Landau and ghost propagators in the position space}

In the momentum space the gluon propagator in Landau gauge is:
\begin{eqnarray}
\dis{\left[g_{\mu\nu} - \frac{p_\mu p_\nu}{p^2}\right] \frac{1}{(p^2)^a},
} 
\ ,
\label{mom}  
\end{eqnarray}
where the case $a=1$ corresponds to the free propagator in four dimensions. 
We will assume that the Wick rotation has been performed and
will thus work in the Euclidean metric. The number of dimensions
is $D = 4 - 2 \e$. We formulate the rules in the momentum space and then we go to the position 
space.  The Fourier transform of the (\ref{mom}) to the position space 
can be done with the help of the following formulas \cite{Kazakov:1984bw}
\begin{eqnarray*}
\dis{\int d^Dp \frac{1}{(p^2)^\a}e^{ipx} = 
2^{D-2\a}\pi^{D/2}a(\a)\frac{1}{(x^2)^{D/2-\a}}, 
~~~~  a(\a) = \frac{\Gamma[D/2-\a]}{\Gamma[\a]}} \ .
 \end{eqnarray*}
Thus, the propagator is of the form 
\begin{eqnarray*}
\dis{\frac{g_{\mu\nu}}{(x^2)^{b}} - c\frac{x_\mu x_\nu}{(x^2)^{b+1}}},   
\end{eqnarray*}
with $b=D/2-a$.

The transversality condition
\begin{eqnarray*}
\dis{\pd_\mu\le \frac{g_{\mu\nu}}{(x^2)^{b}} - 
c\frac{x_\mu x_\nu}{(x^2)^{b+1}} \ri} = 0.   
\end{eqnarray*}
allows to determine the coefficient $c$
\begin{eqnarray*}
& \dis{\pd_\mu\le  \frac{g_{\mu\nu}}{(x^2)^{b}} - 
c\frac{x_\mu x_\nu}{(x^2)^{b+1}} \ri = 
- b \frac{2x_\nu}{(x^2)^{b+1}}- c\frac{(D+1)x_\nu - 
2(b+1)x_\nu}{(x^2)^{b+1}}} = \\ 
& = \dis{-b \frac{2x_\nu}{(x^2)^{b+1}}- c\frac{(D-2b-1)x_\nu}{(x^2)^{b+1}} = 
- \frac{(2b + c(D-2b-1))x_\nu}{(x^2)^{2-\e}}} \\
& \dis{\Rightarrow c = -\frac{2b}{D-2b-1}}.
\end{eqnarray*}
Thus, the free propagator in the Landau gauge is:
\begin{eqnarray*}
\dis{\frac{g_{\mu\nu}}{(x^2)^{1-\e}} + 
2(1-\e)\frac{x_\mu x_\nu}{(x^2)^{2-\e}}}  
\end{eqnarray*}

\vspace{0.5cm}

The ghost propagator in the momentum space is 
\begin{eqnarray}
\dis{\frac{p_\mu}{p^2}} \ .
\end{eqnarray}
In the position space the Fourier transform is 
\begin{eqnarray*}
\dis{\pd_\mu\int d^Dp \frac{1}{p^2}e^{ipx} =   2^{D-2}\pi^{D/2}a(1)\pd_\mu\frac{1}{(x^2)^{1-\e}} =   
2^{D-2}\pi^{D/2}a(1)(\e-1)\frac{2x_\mu}{(x^2)^{2-\e}}  }, 
\end{eqnarray*}

\section{Diagram contributions}

The one-loop contribution to the $Lcc$ correlator corresponds to the diagram of Fig.~1. 
\begin{figure}[ht]
\begin{center}
\epsfig{file=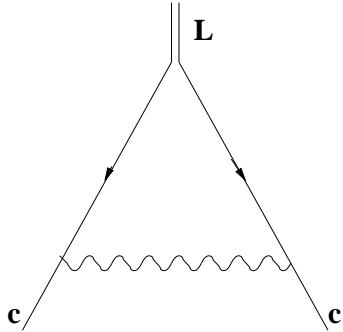, width=5.cm}
\end{center}
\vspace{0.0cm}
\caption{\footnotesize One-loop contribution to the $Lcc$ vertex. The wavy line represents the gluon propagator, the straight lines are for the ghosts.}
\end{figure}
This is the only possible one-loop contribution, since the ghost field interacts only with the gauge field. This one-loop result is simple since it does not 
require the integration in the position space and is proportional to the following expression:   
\begin{eqnarray*}
f^{abc}\int~d^4x_1d^4x_2d^4x_3 L^a(x_1)c^b(x_2)c^c(x_3)\frac{1}{(x_2-x_3)^2}\times\\
\times \le g_{\mu\nu} + 2 \frac{(x_2-x_3)_\mu(x_2-x_3)_\nu}{(x_2-x_3)^2} \ri \pd^{(2)}_{\mu}\pd^{(3)}_{\nu} \frac{1}{(x_1-x_2)^2(x_1-x_3)^2} = \\
2f^{abc}\int~d^4x_1d^4x_2d^4x_3 L^a(x_1)c^b(x_2)c^c(x_3)\left[\frac{1}{(x_1-x_2)^2(x_1-x_3)^4 (x_2-x_3)^2}  \right. \\
\left. + \frac{1}{(x_1-x_2)^4(x_1-x_3)^2 (x_2-x_3)^2} - \frac{2}{(x_1-x_2)^2(x_1-x_3)^2 (x_2-x_3)^4} \right.\\
\left. - \frac{2}{(x_1-x_2)^4(x_1-x_3)^4}  + \frac{1}{(x_1-x_3)^4 (x_2-x_3)^4}  + \frac{1}{(x_1-x_2)^4 (x_2-x_3)^4} \right]
\end{eqnarray*}

As one can see, the one-loop contribution is simple, but it does not have a structure that is expected from conformal field theories 
\footnote{I.K. thanks A. Jevicki for clarifying this point} \cite{Freedman:1998tz,Erdmenger:1996yc}. This is because the external field $L$ does not propagate, 
it is not in the measure of path integral. However, this vertex by Slavnov-Taylor identity can be related to the three-point function of dressed mean gluons
and they are expected to have simple structure at least for the connected function in the Landau gauge in ${\cal N} = 4$ super-Yang-Mills theory. For this reason, it is important 
to calculate the next order of $Lcc$ vertex since poles disappear there. Poles do not disappear in the triple correlator of  gluons since they must be absorbed into the 
dressing functions of the gluons.

\begin{figure}[ht]
\begin{center}
\epsfig{file=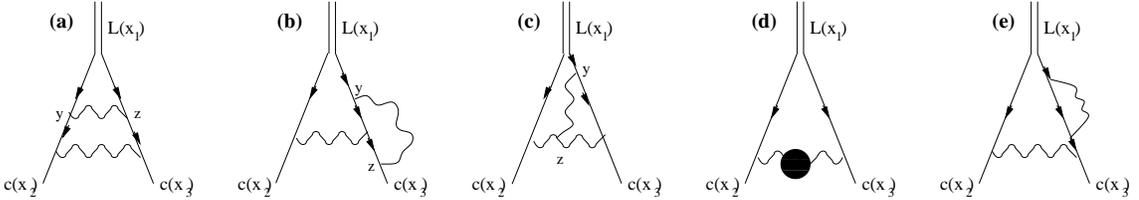, width=15.cm}
\end{center}
\vspace{0.0cm}
\caption{\footnotesize
The two-loop diagrams for the $Lcc$ vertex. The wavy lines represent 
the gluons, the straight lines the ghosts. 
Black disk in diagram (d) stands for the total one-loop correction to the gluon propagator.}
\label{Lccfig}
\end{figure}

Two-loop planar correction  to $Lcc$ vertex can be represented as combination of five diagrams depicted in Fig.~\ref{Lccfig}. 

\section{Integral structure}

As it was noted in the Introduction, in the present paper we analyse diagram $(a)$ only.  
The derivatives on the ghost propagators at the points $x_2$ and $x_3$ can be integrated outside the diagram and be put on the external legs. 
Indeed, the result for the diagram $(a)$ can be represented as 
the derivative
\begin{eqnarray}
\frac{1}{[23]} \le
g_{\mu\nu} + 2 \frac{(23)_\mu(23)_\nu}{[23]} \ri \pd^{(2)}_{\mu}\pd^{(3)}_{\nu}
\label{derivative}
\end{eqnarray}
to an integral $T$ which contains 
numerators of the propagators independent of  $x_2$ and $x_3$. 
We have introduced for the 
brevity the notation 
$$[yz] = (y-z)^2, ~~~~ [y1] = (y-x_1)^2,....$$ 
and so on.

These derivatives will simplify the Lorentz structure of the wavy gluon line and reduce the integrand 
to the scalar structure. The corresponding contribution is of the following form: 
\begin{eqnarray*}
\dis{\frac{(y1)_\mu}{[y1]^{2-\e}}\frac{(z1)_\nu}{[z1]^{2-\e}}
\le\frac{g_{\mu\nu}}{[yz]^{1-\e}} + 
2(1-\e)\frac{(yz)_\mu (yz)_\nu}{[yz]^{2-\e}}\ri\frac{1}{[y2]^{1-\e}}
\frac{1}{[z3]^{1-\e}}} \ . 
\end{eqnarray*} 
Note that  it is the one-loop contribution multiplied by the last two denominators. 

The Lorentz structure can be simplified. Indeed, because the scalar product $2 (xz)_\mu (yz)_\mu = [xy]+[yz]-[xz]$, we have
\begin{eqnarray}
& \dis{\frac{(y1)_\mu}{[y1]^{2-\e}}\frac{(z1)_\nu}{[z1]^{2-\e}}
\le\frac{g_{\mu\nu}}{[yz]^{1-\e}} + 
2(1-\e)\frac{(yz)_\mu (yz)_\nu}{[yz]^{2-\e}}\ri =} \no\\
& \dis{\frac{1}{2}~\frac{ [yz][y1] + [yz][z1]  - (2-\e)[yz]^2  + (1-\e) ( [y1] - [z1])^2 }{[yz]^{2-\e} [y1]^{2-\e} [z1]^{2-\e}}}. \label{sospico}
\end{eqnarray}
 Thus, the integral we have to calculate is   
\begin{eqnarray}
\dis{ T= \int~Dy~Dz~\frac{
 [yz][y1] + [yz][z1]  - (2-\e)[yz]^2  + (1-\e) 
( [y1]^2 + [z1]^2 - 2[y1][z1]) }{ [yz]^{2-\e} [y1]^{2-\e} [z1]^{2-\e}  
[y2]^{1-\e} [z3]^{1-\e}}} 
\ ,
\label{int1}
\end{eqnarray}
where we use the notation $Dy \equiv \pi^{-D/2} d^D y.$ \footnote{we have introduced new $D$-dimensional measure $Dx \equiv  \pi^{-\frac{D}{2}}d^Dx.$ That choice helps us to avoid the 
normalization factor $\pi^{D/2}/(2\pi)^{D}=1/(4\pi)^{D/2}$, coming in calculation of every loop, with the standard measure $d^Dx/(2\pi)^{D}$.The above factor $\pi^{D/2}$ comes from 
integration itself. So, with the new measure $Dx$ the final results in $n$-order of perturbation theory should be added by the factor $\a_s^n/(4\pi)^{Dn/2}$, where $\a_s=g^2_s$ is the 
coupling constant. Thus, in perturbation theory with the expansion parameter $\a_s/(16\pi^2)$, the results in the new measure $Dx$ do not obtain an additional factors in $\overline{MS}$ 
like scheme.}

We expect that the diagram $T$ is finite in the limit $\e \rar 0.$ Infrared divergences in the position space can be analysed in the same manner 
as it has been done for the ultraviolet divergences in the momentum space in the BogolIubov-Parasiuk-Hepp-Zimmermann
$R$-operation. From the expression above, for example, it can be seen that the infrared limit $|x| \rar \infty$ is safe in the position space 
in all the subgraphs and in the whole diagram. In the ultraviolet region of the position space 
each of the integrations is safe, too.

Since the diagram  is finite, {\it does not matter where precisely the ``$\e$'' is.}  In a certain sense, it is possible to change the indices in
the propagators by adding multiples of $\e$. In this way we can achieve the possibility to use the uniqueness relation 
\cite{Vasiliev:1981dg} to calculate at least one of the two integrations by the bootstrap technique. Deviations in logarithms in the integrands, after changing 
the indices, cannot change the results in the limit $\e \to 0$ since they present finite construction times $\e.$  

Thus, with the accuracy $O(\e^0)$, the integral above can be transformed as 
\begin{eqnarray}
\dis{T=
\int~Dy~Dz~\frac{[yz][y1] + [yz][z1]  - (2-\overline{\e})[yz]^2  + 
(1-\overline{\e}) 
( [y1]^2 + [z1]^2 - 2[y1][z1])}{ [yz]^{2} [y1]^{2-2\e} [z1]^{2-2\e} 
[y2] [z3] },}    
\label{int2}
\end{eqnarray}
where $\overline{\e}=-2\e/(1-2\e)$, that corresponds to $c=-2/(1-2\e)$ 
and/or $b=1$. This change $\e \rar \bar{\e}$ is necessary, since we need to keep transversality in the position space 
after changing the index  of gluon propagator from $1-\e$ to 1, according to the  formulas of Section 2. Transversality is necessary requirement to avoid problems with ultraviolet divergence
in the position space.

Moreover, the diagram is symmetric with replacement $\{y,2\} \leftrightarrow 
\{z,3\} $ and,
thus, we replace Eq.~(\ref{int2}) by
\begin{eqnarray}
T= \Biggl\{ 
\int~Dy~Dz~\frac{[yz][z1]  - (1-\overline{\e}/2)[yz]^2  + (1-\overline{\e}) ( [z1]^2 - [y1][z1])}
{ [yz]^{2} [y1]^{2-2\e} [z1]^{2-2\e} [y2] [z3] } \Biggr\} + \no\\
\Biggl\{ \{y,2\} \leftrightarrow \{z,3\} \Biggr\} 
\ .
\label{int2.1}
\end{eqnarray}

Following to the Eq. (\ref{derivative}), the final results for the first diagram $V$ in Fig. 1 can be represented as
\begin{eqnarray}
V ~=~ \frac{1}{[23]} \le
g_{\mu\nu} + 2 \frac{(23)_\mu(23)_\nu}{[23]} \ri \pd^{(2)}_{\mu}\pd^{(3)}_{\nu}
\, T
\label{derivative1}
\end{eqnarray}

\section{Technique of calculation}

To calculate expression (\ref{int2.1}), we need to use uniqueness method \cite{Vasiliev:1981dg,Kazakov:1984km,Kazakov:1984bw}
and Gegenbauer polynomial technique (GPT) \cite{Chetyrkin:1980pr,Kotikov:1995cw}. Let us to give a short review of the uniqueness method. 
The GPT will be used only for the most complicated diagrams in Appendix A. All needed formulas for the GPT application can be found in \cite{Kotikov:1995cw}.

The uniqueness method contains several rules to calculate massless chains and
vertices algebraically, i.e. without a direct calculation of $D$-space
integrals.\\

{\bf 1.}~~ The results for chains $J(\alpha_1,\alpha_2)$  have the form
\begin{eqnarray}
J(\a_1,\a_2)  \equiv \int Dx \frac{1}{[x1]^{\a_1} [x2]^{\a_2}} = A(\a_1,\a_2,\a_3) \frac{1}{[12]^{\tilde{\a}_3}} 
\ , 
\label{Ru1}
\end{eqnarray}
where
\begin{eqnarray*}
 A(\a_1,\a_2,\a_3)= a(\a_1)a(\a_2)a(\a_3), ~~~~  Dx \equiv \pi^{-\frac{D}{2}}d^Dx
\end{eqnarray*}
and
$\a_3=D-\a_2-\a_1$ and $\tilde{\a}_i=D/2-\a_i$. 
The point $x_1$ can be shifted to $x_1=0$. 
We have introduced new $D$-dimensional 
measure $Dx \equiv  \pi^{-\frac{D}{2}}d^Dx.$ Note that chain can be considered as the vertex with one propagator having 
power $0$, i.e. $J(\a_1,\a_2)=J(\a_1,\a_2,0)$.\\

{\bf 2.}~~ {\it Uniqueness method} \cite{Vasiliev:1981dg,Kazakov:1984km}
(see also nice review \cite{Kazakov:1984bw}): if 
$\a_1+\a_2+\a_3=D$, then 
\begin{eqnarray}
J(\a_1,\a_2,\a_3)  \equiv \int Dx \frac{1}{[x1]^{\a_1} [x2]^{\a_2} [x3]^{\a_3}} = A(\a_1,\a_2,\a_3) 
\frac{1}{[12]^{\tilde{\a}_3}[13]^{\tilde{\a}_2}[23]^{\tilde{\a}_1}} 
\ .
\label{Ru2}
\end{eqnarray}

\vskip 0.5cm

{\bf 3.}~~ {\it Integration by parts procedure} (IBP) 
\cite{Vasiliev:1981dg,Kazakov:1984km,Kazakov:1984bw}.\footnote{
In the momentum space the equal relation for triangle is also 
very popular procedure (see \cite{Tkachov:1981wb}).}

Including in the integrand of $J(\a_1,\a_2,\a_3)$ the function
$\pd_{\mu} (x1)_{\mu}$ and applying integration by parts,
we obtain
\begin{eqnarray*}
D \, J(\a_1,\a_2,\a_3)  \equiv \int Dx \frac{1}{[x1]^{\a_1} [x2]^{\a_2}[x3]^{\a_3}} \, \pd_{\mu} (x1)_{\mu} \\
=  \int Dx  \left[ \pd_{\mu} \left\{ \frac{(x1)_{\mu}}{[x1]^{\a_1} [x2]^{\a_2} [x3]^{\a_3}} \right\} -
(x1)_{\mu} \pd_{\mu}\left\{ \frac{1}{[x1]^{\a_1} [x2]^{\a_2} [x3]^{\a_3}} \right\}\right]
\ .
\end{eqnarray*}
The first term on the r.h.s. is equal to zero. Performing the derivative in the second term,  after little algebra we obtain IBP relation
\begin{eqnarray}
\Bigl(D -2\a_1 -\a_2 -\a_3 \Bigr) \, J(\a_1,\a_2,\a_3)
= \a_2 \Bigl( J(\a_1-1,\a_2+1,\a_3) - [12] J(\a_1,\a_2+1,\a_3) \Bigr) 
\nonumber \\
+
\a_3 \Bigl( J(\a_1-1,\a_2,\a_3+1) - [13] J(\a_1,\a_2,\a_3+1) \Bigr)
\ .
\label{Ru3}
\end{eqnarray}

The relation has symmetry with respect to $\a_2 \leftrightarrow \a_3$. The index $\a_1$ has a special role, and the corresponding propagator having the
power $\a_1$ will be called {\it distinguish line}.

\section{Cancellation of poles}

Now, using 
results of Appendices A and B, we obtain the final
results  for 
expression
$T$. For this, it is convenient to consider the
following 
combination:
\begin{eqnarray*}
\overline{T} = \frac{(1-2\e)[12]^{1-\e}[13]^{1-\e}}{\e A(1,1,2-2\e)} \, T ~=~
\sum_{k=1}^{4} \overline{T}_i 
\ .
\end{eqnarray*}

The first term $\overline{T}_1$ has the form
\begin{eqnarray*}
\overline{T}_1 = - \frac{2(1-\e)}{\e} \,
A(1,1,2-2\e),
\end{eqnarray*}
where
\begin{eqnarray*}
A(1,1,2-2\e)= \frac{\Gamma(1+\e)}{\e(1-2\e)} 
\frac{\Gamma^2(1-\e)}{\Gamma(1-2\e)} = \frac{\Gamma(1+\e)}{\e(1-2\e)} 
e^{-\zeta(2)\e^2} + o(\e)
\ .
\end{eqnarray*}
The last identity holds because
\begin{eqnarray}
\Gamma(1+a\e)=\exp \left[-\gamma a\e +\sum_{k=2}^{\infty}\frac{\zeta(k)}{k}
(-a\e)^k \right],
\label{Gamma}
\end{eqnarray}
where $\gamma$ and $\zeta(k)$  are Euler constant and  Euler numbers,
respectively. Thus, we obtain
\begin{eqnarray*}
\overline{T}_1/\{\Gamma(1+\e)e^{-\zeta(2)\e^2}\} = -\frac{2(1-\e)}{\e^2(1-2\e)}  + o(\e).
\end{eqnarray*}

For the second term $\overline{T}_2$ we have
\begin{eqnarray*}
\overline{T}_2 = \frac{2}{(1-2\e)} \, A(\e,2,2-3\e)
\frac{[12]^{\e}[13]^{\e}}{[23]^{2\e}},
\end{eqnarray*}
where
\begin{eqnarray*}
A(\e,2,2-3\e)= -\frac{1-2\e}{2\e(1-3\e)} 
\frac{\Gamma(1-\e)\Gamma(1-2\e)\Gamma(1+2\e)}{\Gamma(1+\e)\Gamma(1-3\e)} = 
-\frac{\Gamma(1+\e)}{2\e} \frac{1-2\e}{1-3\e}  
e^{-\zeta(2)\e^2} + o(\e)
\ .
\end{eqnarray*}
Thus, we obtain
\begin{eqnarray*}
\overline{T}_2/\{\Gamma(1+\e)e^{-\zeta(2)\e^2}\}  = -\frac{1}{\e(1-3\e)} 
\frac{[12]^{\e}[13]^{\e}}{[23]^{2\e}} + o(\e) = 
-\frac{1}{\e(1-3\e)} \Bigl( 1+ \e L \Bigr) + o(\e),
\end{eqnarray*}
where
\begin{eqnarray*}
L=\ln\frac{[12][13]}{[23]^{2}}
\ .
\end{eqnarray*}
 
The third term is very simple
\begin{eqnarray*}
\overline{T}_3 = \Bigl([12]+[13]\Bigr) J(1,1,1),
\end{eqnarray*}
where for integral $ J(1,1,1)$ we can use Davydychev formula 
(see \cite{Davydychev:1992xr}):
\begin{eqnarray}
J(1,1,1) = \frac{2}{B} \Biggl[ \zeta(2) - \Li_2 \left(\frac{[23]+[12]-[13]-B}{2[23]} \right) - \Li_2 \left(\frac{[23]+[13]-[12]-B}{2[23]} \right) \label{Da} \\
+ \ln \left(\frac{[23]+[12]-[13]-B}{2[23]} \right) \ln \left(\frac{[23]+[13]-[12]-B}{2[23]} \right) 
-\frac{1}{2} \ln \left(\frac{[12]}{[23]} \right) \ln \left(\frac{[13]}{[23]} \right) \Biggr]
\ ,
\nonumber
\end{eqnarray}
where
\begin{eqnarray*}
 B^2 = ([12]-[13])^2-2([12]+[13])[23]+[23]^2 
\end{eqnarray*}

Note that 
the results (\ref{Da})
have a clear
symmetry $\{ 2 \leftrightarrow 3 \}$.

The last term  $\overline{T}_4$ has the form
\begin{eqnarray*}
\overline{T}_4 = \frac{2-3\e}{\e} \, 
\biggl[[13]^{1-\e}J(2-3\e,\e,1)+ [12]^{1-\e}J(2-3\e,1,\e) \biggr], 
\end{eqnarray*}
where $J(2-3\e,1,\e)= \Bigl\{J(2-3\e,\e,1), 1 \leftrightarrow 2 \Bigr\}$,
\begin{eqnarray*}
J(2-3\e,\e,1)= \frac{A(1+\e,1,2-3\e)}{A(1+\e,1-\e,2-2\e)} 
\frac{1}{\Gamma(1-\e)(2\e-1)} \,  
\frac{[12]^{-\e}}{[13]^{1-2\e}} \tilde{J}(2-3\e,\e,1),
\end{eqnarray*}
and (see Appendix A)
\begin{eqnarray}
 \tilde{J}(2-3\e,\e,1)=- \frac{1}{\e} + \e \left[ \ln\frac{[12]}{[23]}\ln\frac{[12]}{[23]} + \Bigl([13]+[23]-[12]\Bigr) J(1,1,1) \right]
\ .
\label{tildeJ}
\end{eqnarray}

Because
\begin{eqnarray*}
\frac{A(1+\e,1,2-3\e)}{A(1+\e,1-\e,2-2\e)} = \frac{1-2\e}{2(1-3\e)} 
\frac{\Gamma^2(1-\e)\Gamma(1-2\e)\Gamma(1+3\e)}{\Gamma(1+\e)\Gamma(1-3\e)} = \\
\frac{\Gamma(1+\e)}{2} \frac{1-2\e}{1-3\e}  e^{-\zeta(2)\e^2} + o(\e),
\end{eqnarray*}
we have
\begin{eqnarray*}
\overline{T}_4/\{\Gamma(1+\e)e^{-\zeta(2)\e^2}\} = -\frac{2-3\e}{2\e(1-3\e)} 
\left[\frac{[13]^{\e}}{[12]^{\e}} \tilde{J}(2-3\e,\e,1) +
\frac{[12]^{\e}}{[13]^{\e}} \tilde{J}(2-3\e,1,\e) \right]
+ o(\e).
\end{eqnarray*}

Consider terms in the brackets. All $O(1)$ terms are cancelled:
\begin{eqnarray*}
-\ln\frac{[12]}{[13]} - \ln\frac{[13]}{[12]}=0 
\ .
\end{eqnarray*}

At  
$O(\e)$ level, all logarithms are also canceled. Indeed
 \begin{eqnarray*}
\left\{
-\frac{1}{2} \ln^2\frac{[12]}{[13]} + \ln\frac{[12]}{[23]}\ln\frac{[12]}{[13]}
\right\} + \biggl\{  1 \leftrightarrow 2 \biggr\} = \\
\ln\frac{[12]}{[13]} \left[ - \ln\frac{[12]}{[13]} + \ln\frac{[12]}{[23]}
- \ln\frac{[13]}{[23]} \right] =0 
\ .
\end{eqnarray*}

So, the terms in brackets are
\begin{eqnarray*}
- \frac{2}{\e} + 2\e [23] J(1,1,1) = -2 \left[  
\frac{1}{\e} - \e [23] J(1,1,1) \right]
\ .
\end{eqnarray*}

Thus, we have 
\begin{eqnarray*}
\overline{T}_4/\{\Gamma(1+\e)e^{-\zeta(2)\e^2}\} = \frac{2-3\e}{(1-3\e)}
 \left[  \frac{1}{\e^2} - [23] J(1,1,1) \right]
 + O(\e)
\ ,
\end{eqnarray*}
and, for the sum
\begin{eqnarray*}
\left(\overline{T}_1+\overline{T}_4\right)/\{\Gamma(1+\e)e^{-\zeta(2)\e^2}\} = 
\frac{1}{\e(1-2\e)(1-3\e)} -2 [23] J(1,1,1) + O(\e) =
\left(\overline{T}_1+\overline{T}_4\right)/\Gamma(1+\e). 
\end{eqnarray*}

Combination of three terms
\begin{eqnarray*}
\left(\overline{T}_1+\overline{T}_2+\overline{T}_4\right)/\Gamma(1+\e) = 2 - L - 2[23] J(1,1,1) + O(\e) = \overline{T}_1 + \overline{T}_2 + \overline{T}_4 
\end{eqnarray*}
is finite and, thus,
\begin{eqnarray*}
\overline{T}= 2 - L + \Bigl([12]+[13]-2[23]\Bigr) J(1,1,1) 
\ .
\end{eqnarray*}

Because $\e A(1,1,2-2\e)=1 + o(\e)$, we obtain
\begin{eqnarray}
T= \frac{1}{[12][13]} \biggl[2 - L + \Bigl([12]+[13]-2[23]\Bigr) J(1,1,1) \biggr] 
\ .
\label{diaT}
\end{eqnarray}

\section{Final result}

To obtain the results (\ref{derivative1})
for the diagram $(a)$ of Fig. 1, we apply
the differentiation procedure (\ref{derivative}) to the r.h.s. of Eq. 
(\ref{diaT}).

To write it in the more convenient for the differentiating form it is convenient
to represent the expression in terms of the function $J(1,1,1)$.

It is better to represent the integral $T$ as three terms: 
\begin{eqnarray*}
T=2 T^{(1)} - T^{(2)} + T^{(3)},~~
T^{(1)}= \frac{1}{[12][13]},\\
T^{(2)}= \frac{1}{[12][13]} \, L,~~
 T^{(3)}= \frac{1}{[12][13]} \, \Bigl([12]+[13]-2[23]\Bigr) J(1,1,1).
\end{eqnarray*}

To obtain the final results for the expression $V$ representing the first diagram (a), we apply the
projector 
\begin{eqnarray*}
P_{\mu\nu} \pd^{(2)}_{\mu}\pd^{(3)}_{\nu} \equiv \frac{1}{[23]} \le
g_{\mu\nu} + 2 \frac{(23)_\mu(23)_\nu}{[23]} \ri \pd^{(2)}_{\mu}\pd^{(3)}_{\nu}
\end{eqnarray*}
to the integral $T$, i.e.
\begin{eqnarray*}
V=2 V^{(1)} - V^{(2)} + V^{(3)},~~
V^{(i)} \equiv P_{\mu\nu} \pd^{(2)}_{\mu}\pd^{(3)}_{\nu} T^{(i)}
\ .
\end{eqnarray*}

{\bf 1.}~ The 
result 
for $V^{(1)}$ has the form
\begin{eqnarray*}
V^{(1)}= P_{\mu\nu} \pd^{(2)}_{\mu}\pd^{(3)}_{\nu} \frac{1}{[12][13]}=
\frac{2 B_1}{[12]^2[13]^2[23]^2},
\end{eqnarray*}
where
\begin{eqnarray*}
 B_1= 2 (12)_{\mu}(13)_{\mu}+ 4 (12)_{\mu}(23)_{\mu} (13)_{\nu}(23)_{\nu}
= ([12]-[13])^2+([12]+[13])[23]-2[23]^2 
\ .
\end{eqnarray*}

{\bf 2.}~ To obtain 
$V^{(2)}$, we 
represent it as the sum of three terms:
\begin{eqnarray*}
 V^{(2)}= \sum_{k=1}^{3} V^{(2)}_i 
\ .
\end{eqnarray*}

The first term $V^{(2)}_1$ is proportional to $V^{(1)}$:
\begin{eqnarray*}
V^{(2)}_1= \le P_{\mu\nu} \pd^{(2)}_{\mu}\pd^{(3)}_{\nu} \frac{1}{[12][13]} 
\ri \, L = V^{(1)} \, L
\ .
\end{eqnarray*}

The third term is zero. Indeed
\begin{eqnarray*}
\pd^{(2)}_{\mu}\pd^{(3)}_{\nu} \, L = \pd^{(2)}_{\mu} \left[ -2 
\frac{(13)_{\nu}}{[13]} + 4 \frac{(23)_{\nu}}{[23]} \right] 
= 4 \pd^{(2)}_{\mu} \, \frac{(23)_{\nu}}{[23]} = 4 \left[ - 
\frac{g_{\mu\nu}}{[23]} - 2 \frac{(23)_{\mu}(23)_{\nu}}{[23]^2} \right] \\
\to ~~~
V^{(2)}_3= \frac{1}{[12][13]} P_{\mu\nu} \pd^{(2)}_{\mu}\pd^{(3)}_{\nu} L
= \frac{4}{[12]^2[13]^2[23]^2} \, \Bigl(D+2-2-4\Bigr)=0 .
\end{eqnarray*}

The most complicated second part of $ V^{(2)}$ has the form
\begin{eqnarray*}
V^{(2)}_2= P_{\mu\nu} \left[ \left\{\pd^{(2)}_{\mu} \frac{1}{[12][13]}
\pd^{(3)}_{\nu}  \, L \right\} + \biggl\{  2 \leftrightarrow 3 \biggr\} 
\right] = \tilde{V}^{(2)}_2  + \biggl\{ \tilde{V}^{(2)}_2,  
2 \leftrightarrow 3 \biggr\} 
\ .
\end{eqnarray*}

The result in brackets is
\begin{eqnarray*}
 \frac{2(12)_{\mu}}{[12]^2 [13]}\left[ -2 \frac{(13)_{\nu}}{[13]} + 
 4 \frac{(23)_{\nu}}{[23]} \right] 
\end{eqnarray*}
and, so,
\begin{eqnarray*}
\tilde{V}^{(2)}_2=  \frac{4}{[12]^2[13]^2[23]^2}  \left[
-2 (12)_{\mu}(13)_{\mu}+ 6 (12)_{\mu}(23)_{\mu} -2 (12)_{\mu}(23)_{\mu} 
(13)_{\nu}(23)_{\nu} \right] \\
=  \frac{4}{[12]^2[13]^2[23]^2}  \left[ 2[23]^2- (7[13]+[12])[23]+ 
 ([13]-[12])(5[13]+[12]) \right]
\ .
\end{eqnarray*}

Thus, we have
\begin{eqnarray*}
V^{(2)}_2= \frac{8 B_2}{[12]^2[13]^2[23]^2},~~ 
V^{(2)}= \frac{2 (B_1 L +4B_2) }{[12]^2[13]^2[23]^2},
\end{eqnarray*}
where
\begin{eqnarray*}
 B_2= ([12]-[13])^2- 2([12]+[13])[23]+ [23]^2 \equiv B^2
\ .
\end{eqnarray*}

{\bf 3.}~ The third term $V^{(3)}$ can be also represented as the sum of three 
terms:
\begin{eqnarray*}
 V^{(3)}=  V^{(3)}_1 + V^{(3)}_2 + V^{(3)}_3
\ .
\end{eqnarray*}
where the terms $V^{(3)}_1$, $V^{(3)}_2$ and $V^{(3)}_3$ can be found
in Appendix D.

Collecting all these terms together, we obtain, after some algebra
  \begin{eqnarray*}
V^{(3)} = 
\frac{2}{[12]^2[13]^2[23]^2}
\biggl[ 2 B_4 + 2 B_5 \ln \frac{[12]}{[23]} + 2 B_6 \ln \frac{[13]}{[23]}
+ B_7 J(1,1,1)  \biggr],
\end{eqnarray*}
where
 \begin{eqnarray*}
B_4 &=& \Bigl([13]+[12]\Bigr)^2 -  3\Bigl([13]+[12]\Bigr)[23] + 2 [23]^2, \\
B_5 &=& [12]^2 -[13]^2 +[12][13] + 3[13][23] -2  [23]^2, \\
B_6 &=& \biggl\{B_4,  2 \leftrightarrow 3 \biggr\}, \\
B_7 &=& 4[23]^3 - 6 \Bigl([13]+[12]\Bigr)[23]^2 
+ \biggl(\Bigl([13]+[12]\Bigr)^2 -2 [12][13] \biggr) [23] \\
&& + \biggl(\Bigl([13]+[12]\Bigr)^2 - 6[12][13] \biggr) \Bigl([13]+[12]\Bigr)
\ .
\end{eqnarray*}

{\bf 4.}~ Now the result for the 
expression $V$ 
representing diagram (a)
has the form
  \begin{eqnarray}
V = 2V^{(1)}_1 - V^{(2)} + V^{(3)} =  \frac{2}{[12]^2[13]^2[23]^2}
\biggl[ A_1 + A_2 \ln \frac{[12]}{[23]} + A_3 \ln \frac{[13]}{[23]}
+ A_4 J(1,1,1)  \biggr],
\label{diaV}
\end{eqnarray}
where
 \begin{eqnarray*}
A_1&=&2B_1-4B_2+2B_4 ~=~ 
8[12][13] + 4\Bigl([13]+[12]\Bigr)[23] - 4 [23]^2, \\
A_2&=&-B_1+2B_5 ~=~ [12]^2 -3[13]^2 + 4[12][13] + 
\Bigl(5[13]-[12]\Bigr)[23] -2  [23]^2, \\
A_3 &=& \biggl\{A_2,  2 \leftrightarrow 3 \biggr\}, \\
A_4 &=& B_7 ~=~ 4 [23]^3 - 6 \Bigl([13]+[12]\Bigr)[23]^2 
+ \biggl(\Bigl([13]+[12]\Bigr)^2 -2 [12][13] \biggr) [23] \\
&& + \biggl(\Bigl([13]+[12]\Bigr)^2 - 6[12][13] \biggr) \Bigl([13]+[12]\Bigr)
\ .
\end{eqnarray*}

\section{Conclusions}

In this paper we have shown, among other things, that the first of the five two-loop contributions
in the correlator $Lcc$ does not depend on any scale. The calculation has been performed in the position space
and in the Euclidean metric. For this particular contribution it is a
direct consequence of the transversality of the gluon propagator 
in the Landau gauge. 
The same is true for the 
two other vertex-type contributions. 
The ${\cal N} = 4$ supersymmetry does not play any role in the 
scale-independence of the first three contributions. It is important only 
for the cancellation of poles between the other two contributions of 
propagator-type corrections. The present study was necessary to 
investigate the precise structure of the 
two-loop contributions. By the ST identity the $Lcc$ correlator 
can be transformed to the correlator of three dressed gluons. It is natural to expect that
the conformal symmetry of the theory fixes
the correlator of this triple gluon vertex 
completely up to some coefficient (that depends on the 
gauge coupling and number of colours).
Since it is expected that the structure of the correlators of 
three dressed gluons is simple in the position space, 
the precise structure of the $Lcc$ 
vertex in the position space is also 
simple. 
There are also 
other motivations in favour 
of the expected simple structure of the $Lcc$ vertex.

Indeed, the results for $V$ and $T$ contain the terms with different values of the transcendentality level. It was shown in \cite{Kotikov:2002ab} that
some values of $N=4$ SYM variables have only terms with the same value of the  transcendentality level at 
any order of perturbation theory (see 
\cite{Kotikov:2002ab,Kotikov:2000pm,Bern:2005iz}). 
It is possible that the same property is
applicable to the $Lcc$ vertex. If this is the case, 
then the final two-loop result for the $Lcc$ vertex should 
contain [in the numerator of the 
r.h.s. of Eqs. (\ref{diaT}) and (\ref{diaV})] only $J(1,1,1)$ vertex 
and/or $\zeta(2)$ Euler number.

Finally, we would like to mention another obvious consequence of the considerations made here and in Refs.~\cite{Cvetic:2004kx,Cvetic:2006kk}. 
All this can be applied to any gauge theory with only one coupling (gauge coupling) whose beta function is vanishing 
at every order. It means that the correlators of the dressed gauge bosons in that theory do not depend on any scale 
in the transversal gauge.

\subsection*{Acknowledgments}

The work of I.K. was supported by Ministry of Education (Chile) under grant 
Mecesup FSM9901 and by DGIP UTFSM, by  Fondecyt (Chile) grant \#1040368, and 
by Departamento de Investigaci\'on de la Universidadad del Bio-Bio, Chillan 
(Chile). The work of G.C. and I.S. was supported by Fondecyt (Chile) grants 
\#1050512 and \#1030355, respectively. A.K. was supported by Fondecyt International Cooperation 
Project \#7010094 and 
by Mecesup program of UTFSM. I.K. is grateful to 
Antal Jevicki for discussions of conformal field theories.


\section{Appendix A}
 \label{App:A}
\def\theequation{A\arabic{equation}}
\setcounter{equation}0

The Appendix A is devoted to evaluate the expression for $T$.

The integrand in brackets on the r.h.s. of (\ref{int2.1}) is a sum of several parts. The result, coming from term $\sim [yz]^2$ in numerator, is 
\begin{eqnarray*}
\dis{I_2 ~=~ \int DyDz\frac{1}{[y1]^{2-2\e} [z1]^{2-2\e}  [y2] [z3]}  = 
A^2(1,2-2\e,1) \frac{1}{[12]^{1-\e} [13]^{1-\e}}} 
\ .   
\end{eqnarray*}
This integral is very simple and requires only the use
of Eq. (\ref{Ru1}). The term $\sim [y1][z1]$ leads to
\begin{eqnarray*}
\dis{ I_4 ~=~ \int DyDz\frac{1}{[yz]^2 [y1]^{1-2\e} [z1]^{1-2\e} 
[y2] [z3] } = A(1,2,1 -2\e) \frac{1}{[12]^{-\e}}J(2-3\e,1+\e,1)=} \\
\dis{ A(1,2,1 -2\e) A(2-3\e,1+\e,1) 
\frac{1}{[12]^{1-2\e}[13]^{1-2\e}[23]^{2\e}}},
\end{eqnarray*}
where Eqs. (\ref{Ru1}) and (\ref{Ru2}) have
been used simultaneously. Because of the relations
\begin{eqnarray}
\dis{ A(1,2,1 -2\e) = -(1-2\e) A(1,1,2 -2\e),~~~ A(2-3\e,1+\e,1)=-\frac{1}{1-2\e} A(2-3\e,\e,2), }  \no\\
\dis{ A(1-3\e,1+\e,2) = \frac{2(1-3\e)}{1-2\e} A(2-3\e,\e,2)}
\label{rela}
\end{eqnarray}
we obtain
\begin{eqnarray*}
\dis{ I_4 ~=~  A(1,1,2 -2\e) A(2-3\e,\e,2) \frac{1}{[12]^{1-2\e}[13]^{1-2\e}[23]^{2\e}}}
\ .
\end{eqnarray*}
Using the uniqueness relation (\ref{Ru2}), 
we have for the term $\sim [yz][z1]$
\begin{eqnarray*}
\dis{I_1 ~=~ \int DyDz\frac{1}{[yz] [y1]^{2-2\e} [z1]^{1-2\e}  [y2] 
[z3] } = A(1,1,2 -2\e) \frac{1}{[12]^{1-\e}}J(2-3\e,\e,1)}
\ .
\end{eqnarray*}
The last integral 
\begin{eqnarray*}
\dis{I_3 ~=~ \int DyDz\frac{1}{[yz]^2 [y1]^{2-2\e} [z1]^{-2\e}  [y2] 
[z3] } } 
\end{eqnarray*}
can be reduced by integration by parts to the basic integrals. It is better to consider first the integral $I_1$ and to apply IBP to the vertex with the center in point $z$ and the 
distinguished line with the power $1-2\e$. The result has the form
\begin{eqnarray*}
\dis{ 2\e I_1 ~=~ I_3-I_4 +
\int DyDz\frac{[z1] - [31]}{[yz] [y1]^{2-2\e} [z1]^{1-2\e} 
[y2] [z3]^2 } }
\ . 
\end{eqnarray*}
Evaluating the integrals on the r.h.s., we obtain
\begin{eqnarray*}
\dis{I_3= 
 A(1,1,2 - 2\e)\frac{1}{[12]^{1-\e}} \biggl[ 2\e J(2-3\e,\e,1) - J(1-3\e,\e,2)
\biggr]}  \no\\
\dis{ + \biggl[A(1,2,1-2\e) A(1+\e,1,2-3\e)+ A(1,1,2-2\e) A(\e,2,2-3\e)
\biggr] \frac{1}{[12]^{1-2\e}[13]^{1-2\e}[23]^{2\e}}  } 
\ .
\end{eqnarray*}
Because of the relations (\ref{rela}), the integral $I_3$ has the form
\begin{eqnarray}
\dis{I_3/A(1,1,2-2\e)  = 
\frac{1}{[12]^{1-\e}} \biggl[ 2\e J(2-3\e,\e,1) - J(1-3\e,\e,2)\biggr]}  \no\\ 
\dis{ + 2 A(\e,2,2-3\e)\frac{1}{[12]^{1-2\e}[13]^{1-2\e}[23]^{2\e}}  } 
\ .
\label{I3}
\end{eqnarray}

Both
r.h.s. integrals $J(2-3\e,\e,1)$ and $J(1-3\e,\e,2)$  have 
singularities at $\e \to 0$
and it is convenient to express $J(1-3\e,\e,2)$ through $J(2-3\e,\e,1)$ and 
$J(1-3\e,1+\e,1)$. The last integral is finite at $\e \to 0$.

Applying IBP to the integral $J(1-3\e,\e,2)$ with the 
distinguished
line having 
the power $2$, we 
obtain
\begin{eqnarray*}
\dis{ - J(1-3\e,\e,2) ~=~ \e \biggl[J(1-3\e,1+\e,1)-[23] J(1-3\e,1+\e,2)
\biggr] + (1-3\e) \biggl[J(2-3\e,\e,1) } \no\\
\dis{
-[13] J(2-3\e,\e,2)\biggr] = \e J(1-3\e,1+\e,1) + (1-3\e) J(2-3\e,\e,1) } \no\\
\dis{ - \biggl[\e  A(1+\e,2,1-3\e)
+ (1-3\e) A(\e,2,2-3\e) \biggr]  \frac{1}{[12]^{-\e}[13]^{1-2\e}[23]^{2\e}}  }
\ .
\end{eqnarray*}
Because 
of
relations (\ref{rela}), 
integral $J(1-3\e,\e,2)$ 
obtains
the form
\begin{eqnarray*}
\dis{ - J(1-3\e,\e,2) ~=~ \e J(1-3\e,1+\e,1) + (1-3\e) J(2-3\e,\e,1) } \no\\
\dis{ - \frac{1-3\e}{1-2\e} A(\e,2,2-3\e) 
\frac{1}{[12]^{-\e}[13]^{1-2\e}[23]^{2\e}}  }
\ .
\end{eqnarray*}
Then, for the integral $I_3$ we 
obtain
\begin{eqnarray*}
\dis{I_3/A(1,1,2-2\e)= \frac{1}{[12]^{1-\e}} \biggl[ (1-\e) J(2-3\e,\e,1) +\e J(1-3\e,1+\e,1) \biggr] }  \no\\
\dis{ + \frac{1-\e}{1-2\e} A(\e,2,2-3\e) \frac{1}{[12]^{1-2\e}[13]^{1-2\e}[23]^{2\e}} }
\ .
\end{eqnarray*}

Combining all the results, we obtain for 
expression
$T$:
\begin{eqnarray}
T= I + I(2 \leftrightarrow 3)
\ ,
\label{res}
\end{eqnarray}
where
$I=I_1-(1-\overline{\e}/2)I_2 + (1-\overline{\e}) (I_3-I_4)$ 
has the following 
form:
\begin{eqnarray}
\dis{ \frac{1-2\e}{A(1,1,2-2\e)} \, I = \frac{1}{[12]^{1-\e}} \biggl[ 
(2-3\e) J(2-3\e,\e,1) + 
\e J(1-3\e,1+\e,1) \biggr] }  \no\\
\dis{ +\biggl[ \frac{\e}{1-2\e} A(\e,2,2-3\e) 
\frac{[12]^{\e}[13]^{\e}}{[23]^{2\e}}- 
(1-\e) A(1,1,2-2\e) \biggr]\frac{1}{[12]^{1-\e}[13]^{1-\e}}  }
\ .
\label{res1}
\end{eqnarray}

This formula presents the result for the expression $T$. It contains two integrals  $J(2-3\e,\e,1)$ and $J(1-3\e,1,1+\e)$ which cannot be calculated by rules from the previous 
section and will be calculated in Appendix B by using GPT.

\section{Appendix B}
 \label{App:B}
\def\theequation{B\arabic{equation}}
\setcounter{equation}0

In Appendix B we consider the integral  $J(2-3\e,1,\e)$.\\

{\bf 1.}~~ There is a relation  between two integrals  $J(2-3\e,1,\e)$ and 
$J(\lambda,\lambda,2\lambda)$ where $\lambda = 1 - \e.$ The last integral is 
more convenient to calculate by GPT.

First, we transform using the uniqueness relation

\begin{eqnarray}
\dis{J(2-3\e,\e,1) =  \frac{1}{[23]^{1-2\e}} J(2-3\e,\e,1) [23]^{1-2\e} = 
\int Dx \frac{1}{[23]^{1-2\e} [x1]^{2-3\e} [x2]^{\e} [x3]}[23]^{1-2\e} = }
\no \\
\dis{[23]^{1-2\e}\int Dx  \frac{1}{[x1]^{2-3\e}} \int Dy 
\frac{1}{[yx]^{1+\e} [y2]^{1-\e} [y3]^{2-2\e}}\frac{1}{A(1+\e,1-\e,2-2\e)} =}
\no \\
\dis{[23]^{1-2\e} \frac{A(2- 3\e,1+\e,1)}{A(1+\e,1-\e,2-2\e)}\int Dy 
\frac{1}{[y1]^{1-\e} [y2]^{1-\e} [y3]^{2-2\e}} =} \no \\
\dis{[23]^{1-2\e} 
\frac{A(2- 3\e,1+\e,1)}{A(1+\e,1-\e,2-2\e)}J(1-\e,1-\e,2-2\e)}
.
\label{Ge1}
\end{eqnarray}

{\bf 2.}~~
Thus, we calculate the integral 
appearing above,
by applying  GPT (one can put $x_3 = 0$ by 
shifting the arguments), following \cite{Kotikov:1995cw}
\begin{eqnarray*}
\dis{J(\lambda,\lambda,2\lambda) = \int Dx\sum_{n=0}^{\infty} M_n(\lambda) 
x^{\mu_1\mu_2...\mu_n} x_1^{\mu_1\mu_2...\mu_n}\left[
\frac{\t(x1)}{[x]^{\L + n}} + 
\frac{\t(1x)}{[1]^{\L + n}}  \right]\frac{1}{[x2]^{\L}}\frac{1}{[x]^{2\L}} =} 
\no\\
\dis{\frac{1}{\Gamma(\L)}\sum_{n=0}^{\infty} M_n(\lambda) 
x_1^{\mu_1\mu_2...\mu_n} x_2^{\mu_1\mu_2...\mu_n}\left[
\frac{\t(21)}{[2]^{3\L + n-1}}\frac{1}{(3\L + n-1)(1-2\L)} + 
\right.} \no\\
\dis{\left. \frac{1}{[1]^{3\L + n-1}(n+\L)} \left( 
\frac{\t(12)}{3\L + n-1)} - 
\frac{\t(21)}{1-2\L}\le\frac{[1]}{[2]} \ri^{n+\L} \right) \right.+}\\
\dis{\left. \frac{1}{[2]^{2\L-1}}\frac{\t(12)}{[1]^{\L + n}}
\frac{1}{(2\L -1)(n-\L+1)} - 
\frac{1}{[1]^{3\L+n-1}(n+\L)} \left(
\frac{\t(12)}{2\L -1} -
\frac{\t(21)}{n-\L+1} \le\frac{[1]}{[2]}\ri^{n+\L} \right)
 \right]}
\ .
\end{eqnarray*}

We introduce notation
\begin{eqnarray*}
M_n(\L) \equiv \frac{2^n\Gamma(n+\L)}{n!\Gamma(\L)}, ~~~~ \t(12) 
\equiv \t(x_1^2 -x_2^2)
\ .
\end{eqnarray*}
Here, $M_n(\L)$ is the coefficient at the traceless 
products appearing in the Gegenbauer polynomials 
$C_n^{\L}({\hat{x}_1\hat{x}_2})$
\begin{eqnarray*}
\dis{ C_n^{\L}({\hat{x}_1\hat{x}_2}) = M_n(\L) 
\frac{x_1^{\mu_1\mu_2...\mu_n} x_2^{\mu_1\mu_2...\mu_n}}{(x_1^2)^{n/2}
(x_2^2)^{n/2}} }
\ .
\end{eqnarray*}

Transforming the previous expression one obtains:
\begin{eqnarray*}
\dis{J(\lambda,\lambda,2\lambda) = \frac{1}{\Gamma(\L)} \frac{1}{(1-2\L)}
 \sum_{n=0}^{\infty} M_n(\lambda) x_1^{\mu_1\mu_2...\mu_n} 
x_2^{\mu_1\mu_2...\mu_n}
\left[\t(21)\le\frac{1}{[2]^{3\L + n-1}}\frac{1}{(3\L + n-1)}
- \right.\right.} \no\\
\dis{\left.\left.  -  \frac{1}{[1]^{2\L -1}}\frac{1}{[2]^{n+\L}}
\frac{1}{n-\L+1}\ri
+ \t(12) \le\frac{1}{[1]^{3\L + n-1}}\frac{1}{3\L + n-1} -
 \frac{1}{[2]^{2\L-1}} \frac{1}{[1]^{\L + n}}\frac{1}{n-\L+1} \ri \right] }
\no \\
\dis{\equiv \frac{1}{\Gamma(\L)} \frac{1}{(1-2\L)} 
\biggl[\t(21) j_{21}(\lambda,\lambda,2\lambda) + \t(12) 
j_{12}(\lambda,\lambda,2\lambda)\Biggr]},
\end{eqnarray*}
where $j_{12}(\lambda,\lambda,2\lambda)=
j_{21}(\lambda,\lambda,2\lambda) \Bigl([1] \leftrightarrow [2] \Bigr)$.

Thus, we can consider below only the case $\t(21)$.
It is convenient to use Gegenbauer polynomial itself to reconstruct
the results for $J(\lambda,\lambda,2\lambda)$.

Note that we can re-present the expression for $j_{12}(\lambda,\lambda,2\lambda)$  as   
\begin{eqnarray*}
\dis{j_{21}(\lambda,\lambda,2\lambda) = 
\sum_{n=0}^{\infty} C_n^{\L}({\hat{x}_1\hat{x}_2})
\le\frac{1}{[2]^{2  -3\e}}\frac{\xi^n}{2 + n - 3\e} -  
\frac{1}{[1]^{1-2\e}[2]^{1-\e}}\frac{\xi^n}{n+\e}\ri},
\end{eqnarray*}
where $\xi\equiv \sqrt{[1]/[2]}$. 
We would like to note that GPT has been 
used before only for the
propagator-type diagrams (or for the vertex ones with very specific kinematics --
see the last entry of Refs.~\cite{Chetyrkin:1980pr}), where the results 
could be represented as some numbers, i.e., for example, with
$\xi=1$ and $C^{\lambda}_n(1)$ in above formula. Here, GPT is applied {\it for the 
first time} for the diagrams having two independent arguments. So, we need a 
technique for reconstruction 
of the final results from the expansion in Gegenbauer polynomials. To obtain 
it, we represent the above series in terms of integrals
\begin{eqnarray*}
\dis{ j_{21}(\lambda,\lambda,2\lambda) = } \\
\dis{\sum_{n=0}^{\infty} C_n^{\L}({\hat{x}_1\hat{x}_2}) \le\frac{1}{[2]^{2  -3\e}}\frac{1}{\xi^{2-3\e}} \int_0^{\xi} d\omega \omega^{n+1-3\e} -  
\frac{1}{[1]^{1-2\e}[2]^{1-\e}}\frac{1}{\xi^{\e}}\int_0^{\xi} d\omega \omega^{n+\e-1}\ri =}\\
\dis{\frac{1}{[1]^{1-\frac{3}{2}\e}[2]^{1-\frac{3}{2}\e}}\int_0^{\xi} d\omega \sum_{n=0}^{\infty} C_n^{\L}({\hat{x}_1\hat{x}_2})\le \omega^{n+1-3\e} -  \omega^{n+\e-1}\ri  = }\\
\dis{\frac{1}{[1]^{1-\frac{3}{2}\e}[2]^{1-\frac{3}{2}\e}}\int_0^{\xi} \frac{d\omega \le\omega^{1-3\e}-\omega^{\e-1} \ri}{(1 - 2\cos\t \omega + \omega^2)^{\L}} }
\ .
\end{eqnarray*}
New notation is introduced
\begin{eqnarray*}
({\hat{x}_1\hat{x}_2}) \equiv \cos\t
\ .
\end{eqnarray*}
The second integral over $\omega$ in the above expression  can be re-written as 
\begin{eqnarray*}
\dis{\int_0^{\xi} d\omega \frac{\omega^{\e-1}}{(1 - 2\cos\t \omega + 
\omega^2)^{1-\e}} = \int_0^{\xi} d\omega \omega^{\e}(1 - 2\cos\t \omega + 
\omega^2)^{\e}\left[\frac{1}{\omega} + \frac{2\cos\t - \omega}{1 - 
2\cos\t \omega + \omega^2}\right] =}\\
\dis{\int_0^{\xi} d\omega^{\e}\frac{1}{\e}(1 - 2\cos\t \omega + \omega^2)^{\e} 
+ \int_0^{\xi} d\omega\omega^{\e}\frac{2\cos\t - \omega}{(1 - 2\cos\t \omega 
+ \omega^2)^{1-\e}} = }\\
\dis{\xi^{\e}\frac{1}{\e}(1 - 2\cos\t \xi + \xi^2)^{\e} + \int_0^{\xi} 
d\omega\frac{4\cos\t\omega^{\e} - 3\omega^{1+\e}}{(1 - 2\cos\t \omega + 
\omega^2)^{1-\e}} = }\\
\dis{\frac{1}{\e}\le\frac{[1][12]^2}{[2]^{3}}\ri^{\e/2} + \int_0^{\xi} 
d\omega\frac{4\cos\t\omega^{\e} - 3\omega^{1+\e}}{(1 - 2\cos\t \omega + 
\omega^2)^{1-\e}}}         
\ .
\end{eqnarray*}
In such a way we have extracted the pole in $\e.$ 
The integral in the last line is not singular. 
Now the total expression for $j_{21}$ is 
\begin{eqnarray}
\dis{j_{21}(\lambda,\lambda,2\lambda) = }\no\\
\dis{\frac{[12]^{-\e}}{[1]^{1-2\e}[2]^{1-2\e}} \le
-\frac{1}{\e}\le\frac{[12]}{[2]}\ri^{2\e} + 
\frac{[12]^{\e}}{[1]^{\e/2}[2]^{\e/2}} \int_0^{\xi} d\omega
\frac{-4\cos\t\omega^{\e} + 3\omega^{1+\e} + \omega^{1-3\e}}
{(1 - 2\cos\t \omega + \omega^2)^{1-\e}}  \ri} 
\ .
\label{theta}
\end{eqnarray}
To check the self-consistency of the result, we have to check that the 
theta-functions disappear and that we left with a Lorentz-invariant 
structure. For the poles this is obvious. 
Let us check this for the zeroth order in 
$\e.$  The integral can be decomposed as 
\begin{eqnarray*}
\int_0^{\xi} d\omega\frac{-4\cos\t\omega^{\e} + 3\omega^{1+\e} + 
\omega^{1-3\e}}{(1 - 2\cos\t \omega + \omega^2)^{1-\e}} = 
\int_0^{\xi} d\omega\frac{-4\cos\t + 4\omega }{1 - 2\cos\t \omega + \omega^2}
 + \\
\e\le -4\cos\t \int_0^{\xi} d\omega\frac{\ln\omega}{1 - 2\cos\t \omega + 
\omega^2} + \int_0^{\xi} d\omega\frac{(-4\cos\t + 4\omega)\ln(1 - 
2\cos\t \omega + \omega^2)}{1 - 2\cos\t \omega + \omega^2}\ri + o(\e)
\ .
\end{eqnarray*}
Since we have to calculate the finite part of the initial diagram, 
the higher 
orders in $\e$ do not contribute in the limit $\e \rar 0.$  

The first and the third integral can be easily calculated
\begin{eqnarray*} 
\dis{\int_0^{\xi} d\omega\frac{-4\cos\t + 4\omega }{1 - 2\cos\t \omega + 
\omega^2} =  2\int_0^{\xi} d\ln (1 - 2\cos\t \omega + \omega^2) = 
2\ln (1 - 2\cos\t \xi + \xi^2) = 2\ln\frac{[12]}{[2]}} \ ,
\\
\dis{ \int_0^{\xi} d\omega\frac{(-4\cos\t + 4\omega)\ln(1 - 2\cos\t \omega + 
\omega^2)}{1 - 2\cos\t \omega + \omega^2} = 
\int_0^{\xi} d\ln^2 (1 - 2\cos\t \omega + \omega^2) = \ln^2\frac{[12]}{[2]}}
\ .
\end{eqnarray*}
Moreover, by the  conformal substitution $\omega \rar 1/\omega$ it can 
be immediately verified that   
\begin{eqnarray*} 
\int_0^{\xi} d\omega\frac{\ln\omega}{1 - 2\cos\t \omega + \omega^2} =  
\int_0^{1/\xi} d\omega\frac{\ln\omega}{1 - 2\cos\t \omega + \omega^2},
\ ,
\end{eqnarray*}
i.e. the integral is symmetric 
under the exchange
$[1] \leftrightarrow [2]$.

Expanding the r.h.s. of (\ref{theta}) in powers of $\e$, we obtain
that the terms of zero order in $\e$ are absent
\begin{eqnarray*}
\dis{-2\ln\frac{[12]}{[2]} + 2\ln\frac{[12]}{[2]}=0}
\ ,
\end{eqnarray*}
an that the logarithms of the first order in $\e$ are 
\begin{eqnarray*}
-\ln^2\frac{[12]}{[2]} + \ln \frac{[12]}{[2]}\ln \frac{[12]^2}{[1][2]}
=\ln \frac{[12]}{[2]} \ln \frac{[12]}{[1]}
\ .
\end{eqnarray*}
Thus, we present expression (\ref{theta}) in the form
\begin{eqnarray}
\dis{j_{21}(\lambda,\lambda,2\lambda) = 
\frac{[12]^{-\e}}{[1]^{1-2\e}[2]^{1-2\e}} \left[
-\frac{1}{\e} + \e \le \ln\frac{[12]}{[2]}\ln\frac{[12]}{[1]} 
-4\cos\t \int_0^{\xi} d\omega\frac{\ln\omega}{1 - 2\cos\t \omega + \omega^2} 
\ri \right] } 
\ .
\label{theta1}
\end{eqnarray}

The result (\ref{theta1}) has completely symmetric form under the exchange
$[1] \leftrightarrow [2]$ 
\begin{eqnarray*}
j_{21}(\lambda,\lambda,2\lambda) = j_{12}(\lambda,\lambda,2\lambda)
\end{eqnarray*}
and therefore
\begin{eqnarray}
J(\lambda,\lambda,2\lambda) &=& \frac{1}{\Gamma(1-\e)} \frac{1}{2\e-1}
\frac{[12]^{-\e}}{[1]^{1-2\e}[2]^{1-2\e}} \biggl[
-\frac{1}{\e} \nonumber \\
&&
+ \e \le \ln\frac{[12]}{[2]}\ln\frac{[12]}{[1]} 
-4\cos\t \int_0^{\xi} d\omega\frac{\ln\omega}{1 - 2\cos\t \omega + \omega^2} 
\ri \biggr]  
\ .
\label{9}
\end{eqnarray}

All the theta functions disappeared and the result is 
Lorentz invariant. Now we 
restore $x_3$ component by shifting back
the space-time coordinates $x_1$ and $x_2.$ 
Taking this into account,
we obtain
\begin{eqnarray*}
\cos\t = \frac{(x_1x_2)}{[1]^{1/2}[2]^{1/2}} = 
\frac{1}{2}\frac{[1] + [2] - [12]}{[1]^{1/2}[2]^{1/2}}
\ ,
\end{eqnarray*}
and introducing the notation 
\begin{eqnarray*}
 \int_0^{\xi} d\omega\frac{\ln\omega}{1 - 2\cos\t \omega + \omega^2} \equiv 
I(x_1,x_2,x_3) \equiv I_{123}
\end{eqnarray*}
we can write Eq. (\ref{9}) in the form 
\begin{eqnarray}
\dis{J(\lambda,\lambda,2\lambda) = } \no\\
\dis{\frac{1}{\Gamma(1-\e)} \frac{1}{2\e-1}\frac{[12]^{-\e}}{[13]^{1-\e}[23]^{1-\e}} \left[- \frac{1}{\e} + \e \le \ln\frac{[12]}{[23]}\ln\frac{[12]}{[13]} 
-2 \frac{[13] + [23] - [12]}{[13]^{1/2}[23]^{1/2}}  I_{123} \ri \right] } \label{9.1}
\ .
\end{eqnarray}

{\bf 3.}~~
Now we want to express the integral $I_{123}$ through 
$J(1,1,1)$.
It is convenient to start with the  integral $J(\lambda,\lambda,2\lambda-1)$,
with the purpose to use the results of the previous subsection.

Applying GPT we calculate the integral (one can put $x_3 = 0$ by 
shifting the arguments)
\begin{eqnarray*}
\dis{J(\lambda,\lambda,2\lambda-1) = \int Dx\sum_{n=0}^{\infty} M_n(\lambda) 
x^{\mu_1\mu_2...\mu_n} x_1^{\mu_1\mu_2...\mu_n}\left[
\frac{\t(x1)}{[x]^{\L + n}} + 
\frac{\t(1x)}{[1]^{\L + n}}  \right]\frac{1}{[x2]^{\L}}\frac{1}{[x]^{2\L-1}} 
=} \no\\
\dis{\frac{1}{\Gamma(\L)}\sum_{n=0}^{\infty} M_n(\lambda) 
x_1^{\mu_1\mu_2...\mu_n} x_2^{\mu_1\mu_2...\mu_n}\left[
\frac{\t(21)}{[2]^{3\L + n-2}}\frac{1}{(3\L + n-2)(2-2\L)} + 
\right.} \no\\
\dis{\left. \frac{1}{[1]^{3\L + n-2}(n+\L)} \left( 
\frac{\t(12)}{3\L + n-2} - 
\frac{\t(21)}{2-2\L}\le\frac{[1]}{[2]} \ri^{n+\L} \right) \right.+}\\
\dis{\left. \frac{1}{[2]^{2\L-2}}\frac{\t(12)}{[1]^{\L + n}}
\frac{1}{(2\L -2)(n-\L+2)} - 
\frac{1}{[1]^{3\L+n-2}(n+\L)} \left(
\frac{\t(12)}{2\L -2} -
\frac{\t(21)}{n-\L+2} \le\frac{[1]}{[2]}\ri^{n+\L} \right)
 \right]}
\ .
\end{eqnarray*}

Transforming the previous expression one obtains
\begin{eqnarray*}
\dis{J(\lambda,\lambda,2\lambda-1) =} \\
\dis{ \frac{1}{\Gamma(\L)} \frac{1}{(2-2\L)} \sum_{n=0}^{\infty} M_n(\lambda) x_1^{\mu_1\mu_2...\mu_n} x_2^{\mu_1\mu_2...\mu_n}
\left[\t(21)\le\frac{1}{[2]^{3\L + n-2}}\frac{1}{(3\L + n-2)} - \right.\right.} \no\\
\dis{\left.\left.  -  \frac{1}{[1]^{2\L -2}}\frac{1}{[2]^{n+\L}} \frac{1}{n-\L+2}\ri
+ \t(12) \le\frac{1}{[1]^{3\L + n-2}}\frac{1}{3\L + n-2} - \frac{1}{[2]^{2\L-2}} \frac{1}{[1]^{\L + n}}\frac{1}{n-\L+2} \ri \right] } \no \\
\dis{\equiv \frac{1}{\Gamma(\L)} \frac{1}{(2-2\L)} \biggl[\t(21) j_{21}(\lambda,\lambda,2\lambda-1) + \t(12) j_{12}(\lambda,\lambda,2\lambda-1)\Biggr]},
\end{eqnarray*}
where $j_{12}(\lambda,\lambda,2\lambda-1) = j_{21}(\lambda,\lambda,2\lambda-1) \Bigl([1] \leftrightarrow [2] \Bigr)$. Thus, we can consider below only the case $\t(21)$.
As in the previous subsection, it is convenient to use Gegenbauer polynomial itself to reconstruct the results for $J(\lambda,\lambda,2\lambda)$ obtained above.
Applying the GPT technique one can represent the above expression as   
\begin{eqnarray*}
\dis{j_{21}(\lambda,\lambda,2\lambda-1) = 
\sum_{n=0}^{\infty} C_n^{\L}({\hat{x}_1\hat{x}_2})
\le\frac{1}{[2]^{1  -3\e}}\frac{\xi^n}{1 + n - 3\e} -  
\frac{1}{[1]^{-2\e}[2]^{1-\e}}\frac{\xi^n}{n+1+\e}\ri}.
\end{eqnarray*}
We now use the GTP formulas to present series in terms of integrals 
\begin{eqnarray*}
\dis{j_{21}(\lambda,\lambda,2\lambda-1) = }\\
\dis{\sum_{n=0}^{\infty} C_n^{\L}({\hat{x}_1\hat{x}_2})\le\frac{1}{[2]^{1  -3\e}}\frac{1}{\xi^{1-3\e}} \int_0^{\xi} d\omega \omega^{n-3\e} -  
\frac{1}{[1]^{-2\e}[2]^{1-\e}}\frac{1}{\xi^{1+\e}}\int_0^{\xi} d\omega \omega^{n+\e}\ri =}\\
\dis{\frac{1}{[1]^{\frac{1}{2}-\frac{3}{2}\e}[2]^{\frac{1}{2}-\frac{3}{2}\e}} \int_0^{\xi} d\omega \sum_{n=0}^{\infty} C_n^{\L}({\hat{x}_1\hat{x}_2})\le \omega^{n-3\e} -  
\omega^{n+\e}\ri  =} \\
\dis{\frac{1}{[1]^{\frac{1}{2}-\frac{3}{2}\e}[2]^{\frac{1}{2}-\frac{3}{2}\e}}\int_0^{\xi} d\omega \frac{\le \omega^{-3\e}-\omega^{\e} \ri}{(1 - 2\cos\t \omega + \omega^2)^{\L}} =}\\
\dis{ \frac{1}{[1]^{\frac{1}{2}-\frac{3}{2}\e}[2]^{\frac{1}{2}-\frac{3}{2}\e}}\left[ -4\e \int_0^{\xi} d\omega\frac{\ln\omega}{1 - 2\cos\t \omega + \omega^2} + o(\e) \right] = }\\
\dis{\frac{1}{[1]^{\frac{1}{2}}[2]^{\frac{1}{2}}} \Bigl[ -4\e I_{123} +  o(\e) \Bigr] }
\ .
\end{eqnarray*}
The result has completely symmetric form under $[1] \leftrightarrow [2]$:
\begin{eqnarray*}
j_{21}(\lambda,\lambda,2\lambda-1) = j_{12}(\lambda,\lambda,2\lambda-1)
\end{eqnarray*}
and 
\begin{eqnarray}
\dis{J(\lambda,\lambda,2\lambda-1) = 
\frac{1}{[1]^{\frac{1}{2}}[2]^{\frac{1}{2}}} \Bigl[ -2 I_{123} + o(1) 
\Bigr] } \label{3.1}
\ .
\end{eqnarray}
Now we restore the $x_3$-dependence and obtain the relation 
\begin{eqnarray}
J(1,1,1) = -\frac{2}{[13]^{1/2}[23]^{1/2}}I_{123}
\label{3.2}
\ .
\end{eqnarray}
The following consequences can be derived from here:
\begin{eqnarray*}
\frac{1}{[13]^{1/2}[23]^{1/2}}I_{123} =  
\frac{1}{[12]^{1/2}[23]^{1/2}}I_{132} ~~~~~\Rightarrow  
~~~~~I_{123} =  \frac{[13]^{1/2}}{[12]^{1/2}}I_{132}
\ .
\end{eqnarray*}
Thus, we can write Eq. (\ref{9.1}) in the form 
\begin{eqnarray*}
\dis{J(\lambda,\lambda,2\lambda) = }\\
\dis{\frac{1}{\Gamma(1-\e)} \frac{1}{2\e-1}\frac{[12]^{-\e}}{[13]^{1-2\e}[23]^{1-2\e}} \left[- \frac{1}{\e} + \e \le \ln\frac{[12]}{[23]}\ln\frac{[12]}{[13]} 
+ \Bigl( [13] + [23] -[12] \Bigr) J(1,1,1) \ri \right] } 
\ .
\end{eqnarray*}

\section{Appendix C}
\label{App:C}
\def\theequation{C\arabic{equation}}
\setcounter{equation}0

In Appendix C we consider the integrals $J(1-2\e,2,2)$ and $J(2\e,1,2)$.\\

{\bf 1.}~The integral $J(1-2\e,2,2)$ can be represented as
 \begin{eqnarray*}
J(1-2\e,2,2) = \frac{1}{4\e} \, 
\pd^{(3)}_{\mu}\pd^{(3)}_{\mu} J(1-2\e,2,1)
=\frac{1}{4\e} \, A(1-2\e,2,1) \pd^{(3)}_{\mu}\pd^{(3)}_{\mu}
\frac{[13]^{\e}}{[12]^{1-\e}[23]^{1+\e}}
\ .
\end{eqnarray*}

Because
\begin{eqnarray*}
\pd^{(3)}_{\mu}\pd^{(3)}_{\mu}
\frac{[13]^{\e}}{[12]^{1-\e}[23]^{1+\e}} = \\
8\e (1+\e) \left[ \frac{[13]^{\e}}{[12]^{1-\e}[23]^{2+\e}} - \frac{(23)_{\mu}(13)_{\mu}}{[12]^{1-\e}[13]^{1-\e}[23]^{2+\e}} \right]
+ 4\e \frac{1}{[12]^{1-\e}[13]^{1-\e}[23]^{2+\e}}
\\
= 4\e (1+\e) \left[
\frac{[13]^{\e}}{[12]^{1-\e}[23]^{2+\e}} + \frac{[12]^{\e}}{[13]^{1-\e}[23]^{2+\e}} \right] - 4\e^2 \frac{1}{[12]^{1-\e}[13]^{1-\e}[23]^{2+\e}}
\ ,
\end{eqnarray*}
we have
\begin{eqnarray}
J(1-2\e,2,2)/ A(1-2\e,2,1) = \no\\
(1+\e) \left[\frac{[13]^{\e}}{[12]^{1-\e}[23]^{2+\e}} + \frac{[12]^{\e}}{[13]^{1-\e}[23]^{2+\e}} \right] - \e \frac{1}{[12]^{1-\e}[13]^{1-\e}[23]^{2+\e}}
\ .
\label{B1}
\end{eqnarray}

{\bf 2.}~The integral $J(-2\e,1,2)$ can be rewritten, 
using uniqueness, 
as
\begin{eqnarray*}
\dis{J(-2\e,1,2) =  [12]^{1+\e}
\int Dx \frac{1}{[12]^{1+\e} [x1]^{-2\e} [x2] [x3]^{2}} = }
\no \\
\dis{ [12]^{1+\e} \, \frac{1}{A(2+\e,1-\e,1-2\e)}
\int Dx  \frac{1}{[x3]^{2}}\int Dy 
\frac{1}{[y1]^{1-\e} [y2]^{2+\e} [yx]^{1-2\e}} = }
\no \\
\dis{[12]^{1+\e} \frac{A(2,1- 2\e,1)}{A(2+\e,1-\e,1-2\e)}\int Dy 
\frac{1}{[y1]^{1-\e} [y2]^{2+\e} [y3]^{1-\e}} =} \no \\
\dis{[12]^{1+\e} 
\frac{A(2,1- 2\e,1)}{A(2+\e,1-\e,1-2\e)} J(1-\e,2+\e,1-\e)} 
\ .
\end{eqnarray*}

The integral $J(1-\e,2+\e,1-\e)$ can be treated in a similar manner
\begin{eqnarray*}
\dis{J(2-3\e,\e,1) =  [13]^{\e}
\int Dx \frac{1}{[x1]^{1-\e} [x2]^{2+\e} [x3]^{1-\e}[13]^{\e}} = }
 \\
\dis{ [13]^{\e} \, \frac{1}{A(1,1,2-2\e)}
\int Dx  \frac{1}{[x2]^{2+\e}}\int Dy 
\frac{1}{[y1] [y3] [yx]^{2-2\e}} = }
\\
\dis{[13]^{\e} \frac{A(2+\e,2- 2\e,-\e)}{A(1,1,2-2\e)}\int Dy 
\frac{1}{[y1] [y2]^{2} [y3]} =}  \\
\dis{[13]^{\e} 
 \frac{A(2+\e,2- 2\e,-\e)}{A(1,1,2-2\e)} J(1,2,1)} 
\ .
\end{eqnarray*}

So, for the initial diagram $J(-2\e,1,2)$ we have
\begin{eqnarray*}
\dis{J(-2\e,1,2) =  [12]^{1+\e} [13]^{\e}  J(1,2,1)}, 
\end{eqnarray*}
because
\begin{eqnarray*}
 \frac{A(2,1- 2\e,1)}{A(2+\e,1-\e,1-2\e)} \, 
\frac{A(2+\e,2- 2\e,-\e)}{A(1,1,2-2\e)} = 1
\ .
\end{eqnarray*}

Evaluation of the integral 
$J(1,2,1)$ is very simple. Following Ref.~\cite{Kotikov:1990zk},
we apply IBP to $J(1,1,1)$ with different distinguished lines
\footnote{In momentum space, similar analysis has been done in
\cite{Kotikov:1991hm,Davydychev:1992xr}.}:
\begin{eqnarray}
\dis{ (D-4) J(1,1,1) = 2 J(0,2,1)- [12] J(1,2,1) -[13]J(1,1,2)}, \label{B2} \\
\dis{ (D-4) J(1,1,1) = 2 J(1,0,2)- [23] J(1,1,2) -[12]J(2,1,1)}, \label{B3} \\
\dis{ (D-4) J(1,1,1) = 2 J(1,2,0)- [23] J(1,2,1) -[13]J(2,1,1)}. \label{B4} 
 \end{eqnarray}

Considering the combination 
\begin{eqnarray*}
[23] \cdot <(B2)> - [13] \cdot <(B3)>+ [12] \cdot <(B4)>,
\end{eqnarray*}
where symbols $<(B2)>$, $<(B3)>$ and $<(B4)>$ represent 
the results of Eqs. (\ref{B2}),  (\ref{B3}) and (\ref{B4}), respectively,
we have
\begin{eqnarray*}
\dis{ (D-4) \Bigl([23]-[13]+[12]\Bigr) J(1,1,1) = 
2 \Bigl( [23]J(0,2,1)- [13] J(1,0,2) +[12]J(1,2,0) \Bigr)} \\
\dis{
- 2[23][12] J(1,2,1)}
\ . 
\end{eqnarray*}
The expression on the l.h.s. is negligible when $D \to 4$. Thus
\begin{eqnarray*}
\dis{ J(1,2,1)= \frac{1}{[12][23]} \Bigl( [23]J(0,2,1)- [13] J(1,0,2) 
+[12]J(1,2,0) \Bigr)}
 \end{eqnarray*}
and
\begin{eqnarray}
\dis{ J(1,2,1)/A(1,2,1-2\e) =
\frac{[12]^{\e}[13]^{\e}}{[23]} \biggl[[12]^{-\e}-[13]^{-\e}+[23]^{-\e}
\biggl]}
\ .
\label{B5}
 \end{eqnarray}

\section{Appendix D}
\label{App:D}
\def\theequation{D\arabic{equation}}
\setcounter{equation}0

In Appendix C we calculate the most complicated term $V^{(3)}$ contributed to
the first diagram in Fig.1 .\\

{\bf 1.}~The first term $V^{(3)}_1$ has the form:
\begin{eqnarray*}
V^{(3)}_1/J(1,1,1) = P_{\mu\nu} \pd^{(2)}_{\mu}\pd^{(3)}_{\nu} 
\frac{1}{[12][13]} \Bigl( [12]+[13]-2[23] \Bigr)
\ .
\end{eqnarray*}

The derivatives generate
\begin{eqnarray*}
\pd^{(2)}_{\mu}\pd^{(3)}_{\nu} \frac{1}{[12][13]} \Bigl( [12]+[13]-2[23] \Bigr)
= \frac{4}{[12]^2[13]^2} \biggl( [12][13]  g_{\mu\nu} 
-2 [12](23)_{\mu}(13)_{\nu} \\
+2 [13](23)_{\mu}(12)_{\nu}
-2 [23](12)_{\mu}(13)_{\nu}\biggr).
\end{eqnarray*}

After simple algebra, we have
\begin{eqnarray*}
V^{(3)}_1= \frac{8 B_3}{[12]^2[13]^2[23]}J(1,1,1) ,
\end{eqnarray*}
where
\begin{eqnarray*}
 B_3= ([12]+[13])^2-[12][13] -2 ([12]+[13])[23]+[23]^2 
\ .
\end{eqnarray*}

{\bf 2.}~The third term $V^{(3)}_3$ is
 \begin{eqnarray*}
V^{(3)}_3 \, \frac{[12][13][23]^2}{[12]+[13]-2[23]} = [23]^2 P_{\mu\nu} 
\pd^{(2)}_{\mu}\pd^{(3)}_{\nu} J(1,1,1) =\\
2\le \frac{[23]}{2} \pd^{(2)}_{\mu}\pd^{(3)}_{\mu} J(1,1,1) +
 (23)_{\mu}(23)_{\nu} \pd^{(2)}_{\mu}\pd^{(3)}_{\nu}  J(1,1,1) \ri
\ .
\end{eqnarray*}

The first term in brackets generates the additional factor
 \begin{eqnarray*}
\frac{2(2x)_{\mu}(3x)_{\mu}}{[2x][3x]}=\frac{[2x]+[3x]-[23]}{[2x][3x]}
\end{eqnarray*}
in the  subintegral expression of $J(1,1,1)$, 
where $x$ is the variable of
integration. The  term has the form\footnote{
Hereafter we change $J(1,1,1) \to J(1-2\e,1,1)$ to have regularization and uniqueness of the 
``star''. \cite{Kazakov:1984bw} }

 \begin{eqnarray*}
\frac{1}{2} \pd^{(2)}_{\mu}\pd^{(3)}_{\mu} J(1-2\e,1,1) =
J(1-2\e,1,2)+J(1-2\e,2,1)-[23]J(1-2\e,2,2)
\ .
\end{eqnarray*}

The second term can be calculated similarly: 
$(23)_{\mu}(23)_{\nu} \pd^{(2)}_{\mu}\pd^{(3)}_{\nu} J(1,1,1)$ generates   
 \begin{eqnarray*}
\frac{4(23)_{\mu}(2x)_{\mu}(23)_{\nu}(3x)_{\nu}}{[2x][3x]}
\end{eqnarray*}
in the  subintegral expression of $J(1,1,1)$.

By analogy with above case we obtain, after some algebra 
\begin{eqnarray*}
(23)_{\mu}(23)_{\nu} \pd^{(2)}_{\mu}\pd^{(3)}_{\nu} J(1-2\e,1,1) = 
J(1-2\e,0,2)+J(1-2\e,2,0)- \\
2J(1-2\e,1,1) - [23]^2 J(1-2\e,2,2)
\ .
\end{eqnarray*}
Then
\begin{eqnarray*}
V^{(3)}_3 \, \frac{[12][13][23]^2}{[12]+[13]-2[23]} = \\
2 \biggl[ \biggl\{ [23] J(1-2\e,2,1)+J(1-2\e,2,0)-J(1-\e,1,1) - [23]^2 J(1-2\e,2,2)\biggl\} \\
+ \biggl\{ 2 \leftrightarrow 3 \biggr\} \biggr]~=~2 \biggl[ \tilde{V}^{(3)}_3 + \biggl\{\tilde{V}^{(3)}_3,  2 \leftrightarrow 3 \biggr\} \biggr].
\end{eqnarray*}
Taking the result for $ J(1-2\e,2,2)$ from Appendix B and using
\begin{eqnarray}
J(1-2\e,2,0)=A(1-2\e,2,1) \frac{1}{[12]^{1-\e}} \ , \no\\
J(1-2\e,2,1)=A(1-2\e,2,1) \frac{[13]^{\e}}{[12]^{1-\e}[23]^{1+\e}}
\ ,
\label{resu}
\end{eqnarray}
we have
\begin{eqnarray*}
\Bigl(\tilde{V}^{(3)}_3 +  J(1,1,1)\Bigr)/A(1-2\e,2,1)= \\
\frac{[13]^{\e}}{[12]^{1-\e}[23]^{\e}} +  \frac{1}{[12]^{1-\e}} - (1+\e) \left[\frac{[13]^{\e}}{[12]^{1-\e}[23]^{\e}} + \frac{[12]^{\e}}{[13]^{1-\e}[23]^{\e}} \right] + 
\e \frac{[23]^{1-\e}}{[13]^{1-\e}[12]^{1-\e}} = \\
\frac{1-\e}{[12]^{1-\e}} - \frac{1}{[13]^{1-\e}} \le 1+ \e \left[1+ \ln \frac{[12]}{[23]} \right] \ri + \e \frac{[23]}{[12][13]}
\ .
\end{eqnarray*}
Then 
\begin{eqnarray*}
\Bigl(\tilde{V}^{(3)}_3 + \Bigl\{\tilde{V}^{(3)}_3,  2 \leftrightarrow 3  \Bigr\} + 2 J(1,1,1)\Bigr)/A(1-2\e,2,1) = \\
\e \biggl[ 2 \frac{[23]}{[12][13]}- \frac{1}{[12]} \le 2+ \e \ln \frac{[13]}{[23]} \ri  -  \frac{1}{[13]} \le 2+ \e \ln \frac{[12]}{[23]} \ri \biggr]
\ .
\end{eqnarray*}
Because $\e A(1-2\e,2,1) = -1 + o(\e)$, we have
\begin{eqnarray*}
\tilde{V}^{(3)}_3 + \Bigl\{\tilde{V}^{(3)}_3,  2 \leftrightarrow 3 \Bigr\} +2 J(1,1,1) = \\ 
\frac{1}{[12][13]}\biggl[ 2 \le [12]+[13]-[23] \ri +[12] \ln \frac{[12]}{[23]} + [13] \ln \frac{[13]}{[23]}  \biggr]
\ .
\end{eqnarray*}
Thus, 
\begin{eqnarray*}
V^{(3)}_3 =  \frac{2([12]+[13]-2[23])}{[12]^2[13]^2[23]^2}\biggl[ 2 \le [12]+[13]-[23] \ri +[12] \ln \frac{[12]}{[23]}  + [13] \ln \frac{[13]}{[23]}\biggr.\\ 
\biggl.  - 2 [12][13] J(1,1,1) \biggr]
\end{eqnarray*}

{\bf 3.}~
The most complicated second part of $ V^{(3)}$ has the form
\begin{eqnarray*}
V^{(3)}_2= P_{\mu\nu} \left[ \left\{\pd^{(2)}_{\mu} \le \frac{1}{[12][13]}
 \le [12]+[13]-2 [23] \ri \ri
\pd^{(3)}_{\nu}  \, J(1,1,1)  \right\} + 
\biggl\{  2 \leftrightarrow 3 \biggr\} \right] 
\ .
\end{eqnarray*}

The derivative $\pd^{(2)}_{\mu}$ generates
\begin{eqnarray*}
\pd^{(2)}_{\mu} \le \frac{1}{[12][13]}
 \le [12]+[13]-2 [23] \ri \ri
=  \frac{2}{[12]^2[13]} \left[ \Bigl([13]-2[23]\Bigr) (12)_{\mu}
- 2[12](23)_{\mu} \right]
\end{eqnarray*}
and after little algebra we have
\begin{eqnarray*}
P_{\mu\nu} \pd^{(2)}_{\mu} \le \frac{1}{[12][13]}
 \le [12]+[13]-2 [23] \ri \ri
=  \frac{2}{[12]^2[13][23]^2} \left[ \Phi_1 (12)_{\nu}
+ \tilde{\Phi}_2 (23)_{\nu} \right],
\end{eqnarray*}
where
\begin{eqnarray*}
 \Phi_1 =  [23]\Bigl([13]-2[23]\Bigr),~~
 \tilde{\Phi}_2 =  [13]\Bigl([13]-[12]\Bigr) - [23]\Bigl(3[13]+4[12]\Bigr)
+2  [23]^2 .
\end{eqnarray*}

It is convenient to represent $(12)_{\nu}$ as 
$(12)_{\nu}=(13)_{\nu}-(23)_{\nu} $. Then
\begin{eqnarray*}
P_{\mu\nu} \pd^{(2)}_{\mu} \le \frac{1}{[12][13]}
 \le [12]+[13]-2 [23] \ri \ri
=  \frac{2}{[12]^2[13][23]^2} \left[ \Phi_1 (13)_{\nu}
+ \Phi_2 (23)_{\nu} \right],
\end{eqnarray*}
where
\begin{eqnarray*}
\Phi_2 =  [13]\Bigl([13]-[12]\Bigr) - 4\Bigl([13]+[12]\Bigr)[23]
+4  [23]^2 .
\end{eqnarray*}

Thus, the considered term $V^{(3)}_2$ has the form
 \begin{eqnarray*}
V^{(3)}_2 \, {[12]^2[13]^2[23]^2} = 
2 \biggl[ \biggl\{ [13] \Bigl[\Phi_1 (13)_{\nu} 
+ \Phi_2 (23)_{\nu} \Bigr] \pd^{(3)}_{\nu}  \, J(1,1,1) \biggl\} 
+ \biggl\{ 2 \leftrightarrow 3 \biggr\} \biggr] \\
~=~
2 \biggl[ \tilde{V}^{(3)}_2 + \biggl\{\tilde{V}^{(3)}_2,  2 \leftrightarrow 3 
\biggr\} \biggr],
\end{eqnarray*}
where
 \begin{eqnarray*}
 \tilde{V}^{(3)}_2 = [13] \Bigl[\Phi_1 W^{(3)}_1
+ \Phi_2 W^{(3)}_2 \Bigr]
\ .
\end{eqnarray*}

Consider 
first
$W^{(3)}_2= (23)_{\nu}\pd^{(3)}_{\nu}  \, J(1,1,1)$.
The derivative generates the term
 \begin{eqnarray*}
\frac{2(23)_{\nu}(3x)_{\nu}}{[3x]}=-\frac{[23]+[3x]-[2x]}{[3x]}
\end{eqnarray*}
in the  subintegral expression of $J(1,1,1)$.

Thus, $W^{(3)}_2$ is
 \begin{eqnarray*}
W^{(3)}_2 -  J(1,1,1)= - J(1-2\e,0,2) + [23] J(1-2\e,1,2)
\ .
\end{eqnarray*}

Using Eq.(\ref{resu}) we obtain
 \begin{eqnarray*}
\Bigl(W^{(3)}_2 -  J(1,1,1)\Bigr)/A(1-2\e,2,1)= 
\frac{[12]^{\e}}{[13]^{1-\e}[23]^{\e}}  - \frac{1}{[13]^{1-\e}}
=   \frac{\e}{[13]} \ln \frac{[12]}{[23]} 
\ .
\end{eqnarray*}

Because $\e A(1-2\e,2,1) = -1 + o(\e)$, we have
 \begin{eqnarray*}
W^{(3)}_2 = - \frac{1}{[13]} \ln \frac{[12]}{[23]} +  J(1,1,1)
\ .
\end{eqnarray*}

The operation
$W^{(3)}_1= (13)_{\nu}\pd^{(3)}_{\nu} \, J(1,1,1) $
 generates the term
 \begin{eqnarray*}
\frac{2(13)_{\nu}(3x)_{\nu}}{[3x]}=-\frac{[13]+[3x]-[1x]}{[3x]}
\end{eqnarray*}
in the  subintegral expression of $J(1,1,1)$.

Thus, $W^{(3)}_1$ is
 \begin{eqnarray*}
W^{(3)}_1 -  J(1,1,1)= - J(2\e,1,2) + [13] J(1-2\e,1,2)
\ .
\end{eqnarray*}
Taking the integral $J(2\e,1,2)$ from Appendix B and
using Eq.(\ref{resu}), we obtain
\begin{eqnarray*}
\Bigl(W^{(3)}_1 -  J(1,1,1)\Bigr)/A(1-2\e,2,1) = \\
\frac{[12]^{\e}[13]^{\e}}{[23]} \biggl[ \frac{1}{[23]^{\e}} - \Bigl(\frac{1}{[12]^{\e}}-\frac{1}{[13]^{\e}}+\frac{1}{[23]^{\e}} \Bigr)\biggr] =   \frac{\e}{[23]} \ln \frac{[12]}{[13]} 
\ .
\end{eqnarray*}
Thus,
 \begin{eqnarray*}
W^{(3)}_1 = - \frac{1}{[23]} \ln \frac{[12]}{[13]} +  J(1,1,1).
\end{eqnarray*}
Then
\begin{eqnarray*}
\tilde{V}^{(3)}_2 =  - \frac{[13]}{[23]} \Phi_1 \ln \frac{[12]}{[13]} - \Phi_2 \ln \frac{[12]}{[23]} +[13] \Bigl(\Phi_1 + \Phi_2) J(1,1,1)
\end{eqnarray*}
and
 \begin{eqnarray*}
V^{(3)}_2 =  \frac{2}{[12]^2[13]^2[23]^2}
\biggl[ \phi_1 \ln \frac{[12]}{[23]} + \phi_2 \ln \frac{[13]}{[23]}
+ \phi_3 J(1,1,1)  \biggr],
\end{eqnarray*}
where
 \begin{eqnarray*}
 \phi_1 &=& - \Phi_2 + \Bigl([12]-[13]\Bigr) \Bigl([13]+[12]-2[23]\Bigr)
=  \Bigl([12]-[13]\Bigr)\Bigl(2[13]+[12]\Bigr) \\
&& + 2\Bigl([12]+3[13]\Bigr)[23] -4  [23]^2, \\
 \phi_2 &=& \biggl\{\phi_1,  2 \leftrightarrow 3 \biggr\}, \\
\phi_3 &=& [13] \Bigl(\Phi_1+\Phi_2 \Bigr) + [12]
\biggl\{ \Bigl( \Phi_1+\Phi_2 \Bigr),  2 \leftrightarrow 3 \biggr\}
= \Bigl([13]+[12]\Bigr) \biggl(\Bigl([13]-[12]\Bigr)^2 + 2[23]^2 \biggr)\\
&& - \biggl(3\Bigl([13]+[12]\Bigr)^2 + 2[12][13] \biggr) [23]
\ .
\end{eqnarray*}

\end{document}